      \documentclass[aps,prl,twocolumn,showpacs]{revtex4}
  \usepackage{graphicx}%

\begin{document}
\title{Correlation lengths of  Wigner crystal order in
two dimensional electron system at high magnetic field 
}

\author{P.D. Ye$^{1,2}$, L.W. Engel$^1$, D.C. Tsui$^2$, R. M. Lewis$^{1,2}$, L.N. Pfeiffer$^{3}$, K. West$^{3}$} 
\affiliation{$^1$National High Magnetic Field Laboratory, Tallahassee, 
Florida 32310}
\affiliation{$^2$Department of Electrical Engineering, Princeton University, 
Princeton, New Jersey 08544}
\affiliation{$^3$Bell Labs, Lucent Technologies, Murray Hill, NJ 07974}

\begin{abstract}    
The insulator terminating the fractional quantum Hall series at low 
Landau level filling $\nu$ is generally taken to be  a pinned 
Wigner crystal (WC), and exhibits a microwave resonance that is interpreted
as a WC pinning mode. Systematically studying the resonance in a high 
quality sample for carrier densities, $n$, between  1.8  and 5.7 $\times 
10^{10}$ cm$^{-2}$, we find maxima in 
resonance peak frequency, \mbox{$f_{pk}$}, vs  magnetic field, $B$.  $L$,
the domain size, or correlation length of  
Wigner crystalline order,  is calculated from \mbox{$f_{pk}$}.  For each 
$n$,  $L$ vs $\nu$    tends at low $\nu$ toward a  straight line with 
intercept; the linear fit is  accurate over as much 
as a factor of 5  range of $\nu$.   We interpret the  striking linear behavior 
as due to $B$ compressing the electron wavefunctions, 
to alter the effective electron-impurity interaction.   
\end{abstract}
 
\maketitle

 The nature of a two dimensional electron system (2DES)  in high magnetic 
field ($B$) depends on    the interactions of electrons with each other 
and with impurities,  and on  the overlap of individual 
electron wavefunctions. 
The size of an electron wavefunction, as measured by the 
magnetic length, $l_B=\sqrt{ \hbar/eB}$, controls this overlap, and 
can also affect the electron-impurity interaction if the effective 
size or spacing of the impurities is comparable to $l_B$.   
In the high $B$ limit, $l_B$ vanishes, and  electrons look like 
classical point particles. 
Without disorder, they are expected to form a Wigner crystal
(WC), which is a triangular lattice stabilized by interelectron 
repulsion.  Introduction of small amounts of disorder   pins the WC, making it an  
insulator, and causing the crystalline order to have a finite correlation length
or domain size. 
The   wavefunction overlap increases as 
$B$ is decreased from the high $B$ limit; its importance is 
characterized by $l_B/a$, where $a$ is the WC lattice constant, or 
equivalently by the Landau filling factor 
$\nu=nh/eB=(4\pi/\sqrt{3})(l_B/a)^2$.  
At sufficiently high  $\nu$,
 calculated \cite{wcpredict,kunwc} for disorder free systems 
to be  around 1/7, the WC ground state   is predicted to undergo a
transition to  the  fractional quantum Hall 
effect (FQHE)\cite{fqheorig} liquids.

Experimentally, 2DES are insulators
   from the maximum $B$ accessed down 
at least to the high $B$ edge of the 1/5 FQHE plateau\cite{msreview,reentrant}. 
Samples of sufficient quality to exhibit  the  1/3 
FQHE  have    a well-defined 
resonance\cite{williams,m124,clibdep,clidensity} in the microwave 
spectrum of  the high $B$  insulating phase.  The natural 
interpretation of the resonance is as a pinning mode of the WC, in 
which  regions of WC oscillate within the impurity potential that 
pins  them.  The average (per electron) restoring force constant, $K$,   on static 
displacement away from equilbrium positions 
determines the frequency of the resonance peak.    This force 
constant is $K=m^* \omega_0^2$, which defines $\omega_0$, 
the pinning frequency\cite{flr,nlm}, with $m^*$ the carrier effective  mass.
In the magnetic field, cyclotron motion mixes with the oscillation 
in the pinning potential for the resonating 2DES, resulting\cite{flr}  in the observed 
resonance peak frequency $f_{pk}=\omega_0^2/2\pi\omega_c$, as long as
the cyclotron frequency $\omega_c\ll \omega_0$, which must be the case 
for resonances in the frequency range of interest.  In the classical, 
high $B$ limit, in which $l_B$ is much smaller than any feature of the 
disorder, and  also small enough that wavefunction overlap of neighboring 
electrons can be neglected, $\omega_0$ is constant in $B$, and 
$f_{pk}\propto 1/B$.      

Correlation lengths of crystalline order, or domain sizes,  can 
be calculated directly from \mbox{$f_{pk}$} and  
the elastic constants of the WC.  Recent 
theories\cite{fertig,chitra,fogler} identify the length 
directly relevant  to $f_{pk}$    
as the Larkin length, $L$, over which  
displacements of equilibrium positions of electrons
from perfect crystalline  order reach  the characteristic 
length  
of the electon-impurity interaction.  In the classical, 
high $B$ limit, the dependence of $L$ on $B$ has saturated, and $L$ is 
a constant.      Earlier theory\cite{flr,nlm} considered  the 
particular case of a sinusoidal charge density wave in magnetic field, 
and so obtained $f_{pk}$ in terms 
of a correlation length that we denote $L_a$, over which  displacements 
of equilibrium positions of electrons
from perfect crystalline  order reaches the lattice constant $a$.

This paper  presents data on the microwave resonance for a high 
quality 2DES, surveyed over a wide range of $B$ up to 20 T,  and 
density, $n$ between 5.7 and 1.8 $\times 10^{10} $ cm$^{-2}$ , to cover $\nu$ 
from 0.2 to 0.038.   For all the  $n$'s we looked at,  the resonance peak frequency, 
$f_{pk}$, vs $B$  exhibits a maximum.
Most likely because of higher quality in the present sample, such a 
maximum was not found in previous studies\cite{williams,m124,clibdep,clidensity},  
in which  \mbox{$f_{pk}$} vs 
$B$ was increasing monotonically.   We 
calculate $L$ from \mbox{$f_{pk}$} and the  theoretical shear modulus of a WC of classical 
point particles, and find   $L$ in the range of 0.23 to 0.71 $\mu$m. 
Plots of $L$ vs $\nu$ (with $n$ fixed) tend toward a 
straight line with intercept, $L_\infty$, at low $\nu$, with upward 
curvature developing at most $n$'s as $\nu=1/5 $ is approached from below.  
The linear form for the $L$ vs $\nu$ data is quantitatively  
convincing for the lowest few densities surveyed; at the lowest $n$
there is a precise fit to the data   over a  factor of 5 in $\nu$.   
The large ranges of linear fit  for low $n$, extending to such low 
$\nu$, suggest  the linear behavior is an effect of the interplay of 
$l_B$ and impurity correlation lengths.  We   interpret the 
upward curvature that 
sets in at higher $\nu$  as an effect of the correlations 
responsible for the FQHE.


  The microwave measurement methods used here are similar to those 
described in earlier publications \cite{clibdep,clidensity}.
The inset to Fig. 1 shows a  sketch of the sample. 
Metal film, shown as black in the figure, was patterned onto the top 
of the sample to  
form a coplanar transmission line  consisting of a 45 $\mu$m wide center 
strip separated from side planes by slots  of width $W=30$ $\mu$m.  
The narrow center conductor 
was driven and the two side conductors   
were grounded, and the line 
coupled capacitively to the 2DES.   The geometry confines the microwave 
field mainly to the regions of 2DES under the slots.  

The transmitted
microwave power was measured and normalized to unity for the case of 
vanishing $\sigma_{xx}$, obtained by depleting the 2DES with backgate voltage.  
The experiment is sensitive to $\sigma_{xx}$ with
wavevector $q\lesssim 2\pi/W$; we regard the 2DES as  
in the $q=0$ limit in calculating  the real part of 
diagonal conductivity, \mbox{Re$(\sigma_{xx} )$}, from the 
microwave absorption.    
We  used    Re$(\sigma_{xx})=W|\ln P|/2Z_0d$, 
where $P$ is the normalized transmitted power, $Z_0=50 \Omega$, the  $\sigma_{xx}=0$
characteristic impedance, and $d=28$ mm is 
the total length of the transmission line.  This formula is valid in 
the high $f$, low loss limit, in the absence of reflections. 
Detailed simulation  of the  transmission line  with 2DES 
 in the quasi-TEM approximation   indicates 
  this formula is correct to about  15\% under 
experimental conditions.   
The  apparatus is typically 20 times more sensitive to 
 Re$(\sigma_{xx})$ than it is to Im$(\sigma_{xx})$. 
 The temperature   of  
all measurements presented here was $\sim$ 50 mK. Microwave power was 
varied to ensure that the measurement was in a linear regime at the 
operating temperature. 

   The data were taken on a  
AlGaAs/GaAs heterojunction sample with  as-cooled  
density ($n$) of $6.0\times 10^{10}$ cm$^{-2}$ and   0.3 K mobility of $6 
\times 10^6$ cm$^2$/V s.   The sample was not illuminated, and its 
density was varied down to $1.8\times 10^{10}$ cm$^{-2}$ by 
application of a voltage between a backgate and an ohmic contact, 
  placed outside the transmission line film.

\begin{figure}[t]    
   \includegraphics[scale=0.44]{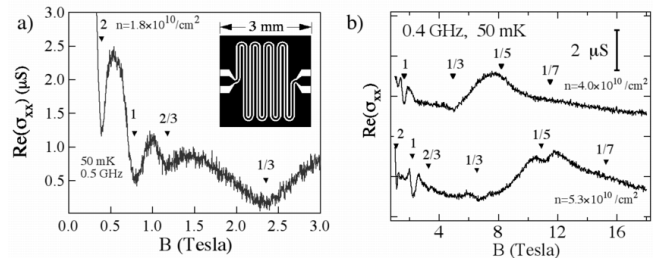} 
    
 \caption{ Real part of diagonal conductivity \mbox{Re$(\sigma_{xx} )$} vs magnetic 
 field, $B$, for the sample at three different carrier densities $n$, 
 filling factors $\nu$ marked as inverted triangles. 
 {\em a.\ \ } $n= 1.8 \times 10^{10}$ cm$^{-2}$. Inset shows top view of 
 transmission line, where black indicates metal film. 
 {\em b.\ \ } $n=4.0 \times 10^{10}$ cm$^{-2}$ and 
   $n=5.3 \times 10^{10}$ cm$^{-2}$.\vspace*{-10pt} }
 \end{figure}
 
Figure 1 shows  traces of \mbox{Re$(\sigma_{xx} )$} vs $B$  in the quantum Hall regime
for three different  $n$'s, for 
$f=0.4$ or 0.5 GHz.  The 1/3 FQHE is present for all $n$ we survey; 
even the lowest $n$, $1.8\times 10^{10}$ cm$^{-2}$, in Figure 1a, exhibits well 
defined FQH features, indicating the 2DES remains reasonably 
homogenous even at the large backgate bias required to produce that $n$.
The 1/5 FQHE, however,  is only present for    
larger $n$, as seen in Figure 1b,  where a 1/5 FQHE minimum in 
\mbox{Re$(\sigma_{xx} )$} vs $B$ is present for $n\approx 5.3 \times 10^{10}$ cm$^{-2}$ but 
absent for  $4.0\times 10^{10}$ cm$^{-2}$.    The  
traces of Figure 1b show broad peaks beginning just above the 1/3 
FQHE. These peaks are the manifestations in \mbox{Re$(\sigma_{xx} )$} of the rapidly rising 
 dc $\rho_{xx}$ vs $B$   \cite{reentrant,sajoto}, observed on  increasing 
 $B$  beyond the 1/3 FQHE. The 1/5 FQHE minimum   for $n\approx 4.0\times 10^{10}$ cm$^{-2}$
is superimposed on that peak.

 
\begin{figure*}[t] 
\includegraphics[scale=0.85]{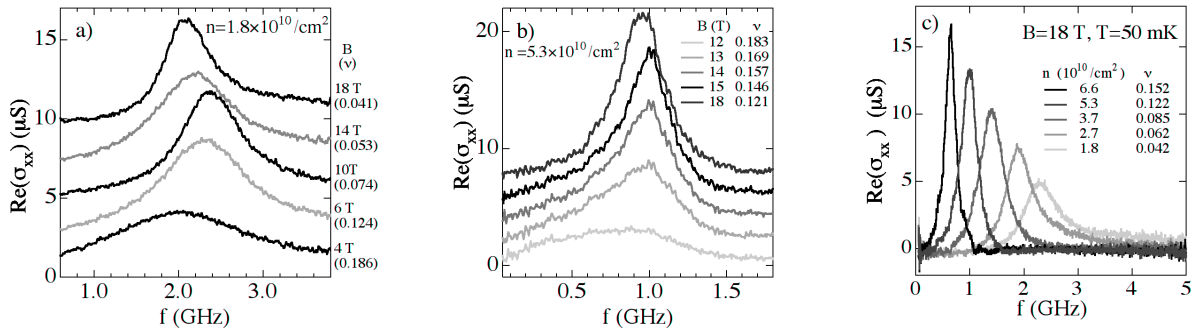} 
\caption{ Resonance spectra, real part of
diagonal conductivity, \mbox{Re$(\sigma_{xx} )$}, vs frequency $f$.  {\em a.\ \ }
Density $n=1.8 \times 10^{10}$ cm$^{-2}$, various magnetic fields $B$.
Successive curves offset in steps of 2.5 $\mu$S, for clarity; the 4 T 
curve is not offset.  
{\em b.\ \ }
Density $n=5.3 \times 10^{10}$ cm$^{-2}$, various   $B$.
Successive curves offset in steps of 2.0 $\mu$S, for clarity. 
{\em c.\ \ } Various $n$, for fixed $B=18 $ T. }
 \end{figure*}

Figure 2 shows typical resonance spectra, \mbox{Re$(\sigma_{xx} )$} vs $f$, taken at 
various $n$ and $B$.  Spectra for several $B$ are shown in Figure~2a for 
$n=1.8\times 10^{10}$ cm$^{-2}$ and in Figure ~2b for $n=5.3\times 
10^{10}$ cm$^{-2}$.  For $\nu$ just below 1/5, the resonance is 
present but comparatively broad; increasing $B$ initially sharpens the 
peak and shifts it to higher $f$.  At some $n$-dependent magnetic 
field,
the resonance peak frequency, \mbox{$f_{pk}$}, 
goes through a maximum and begins to shift {\em downward}  as $B$ is 
increased further.   The reduction of \mbox{$f_{pk}$} continues, 
and the resonance remains well-developed, out to the maximum $B$ 
measured.   The maximum \mbox{$f_{pk}$} occurs around 10 T ($\nu=0.074$) 
for the $n=1.8\times 10^{10}$ cm$^{-2}$and around 15 T ($\nu=0.146$) 
for $n=5.3\times 10^{10}$ cm$^{-2}$. 
  
Figure~2c shows 
resonance spectra for fixed $B=18 $ T as $n$ is varied.  The 
resonances shift to higher $f$ and broaden as $n$ is decreased.  
  Upward shift and broadening  as $n$ is decreased at fixed $B$  
is seen for all $B$ where the resonance is observed, and has 
been observed in other samples\cite{clidensity,williams} as 
well.
 
Figure~3a shows \mbox{$f_{pk}$} vs $\nu$ for the various $n$'s we studied.  
Each curve exhibits a maximum in \mbox{$f_{pk}$}, and the maxima shift generally 
to lower $\nu$ as $n$ decreases.  
 Full width at half maximum linewidths, $\Delta f$, are
plotted against the same axis in Figure~3b.   
These increase rapidly  as the transition to the FQH liquid is 
approached.  Around the  $\nu $  where \mbox{$f_{pk}$} is maximal,  the 
corresponding $\Delta f$ curve for the same $n$ shows a local minimum.  

 %

 The  increasing   \mbox{$f_{pk}$} vs $\nu$  curves (for fixed $n$) on the low $\nu$ side of the maxima 
may be viewed only as  the  approach to the classical  regime in which 
$\mbox{$f_{pk}$}\propto \nu$ is expected.   The observed increase  is  sublinear, and 
while the increase 
 of \mbox{$f_{pk}$} with $\nu$ is a hint toward the classical behavior, 
 the fully classical,  point particle  picture  is not  
applicable.  
The decreasing \mbox{$f_{pk}$} vs $\nu$  (increasing \mbox{$f_{pk}$} vs $B$),  seen in the present data set on the high
$\nu$ side of the maxima, and throughout previous data 
sets\cite{m124,clibdep},  has been  interpreted in theories\cite{fertig,chitra,fogler}  
based on an interplay of  $l_B$ and some disorder correlation length, 
$\xi$, whose definition   depends on the model of disorder.
The calculations,  taken for weak disorder with small $\xi\ll l_B$, 
  give\cite{fertig}   $ \mbox{$f_{pk}$}\propto  B$, or\cite{chitra,fogler} 
$  \mbox{$f_{pk}$}\propto  B^2$ .

To more simply interpret the  behavior of \mbox{$f_{pk}$} vs $\nu$ 
we recast  the  \mbox{$f_{pk}$}  data  into  domain size, $L$,
using  the   elastic coefficients of the  classical WC. 
$L$ is the localization length of transverse phonons 
in the $B=0$ crystal\cite{fogler,bzeronote}.  
The angular pinning frequency, $\omega_0=c_t 2\pi /L$, where $c_t$ is the $B=0$  transverse
phonon progagation velocity,  $c_t= (\mu_t/nm^*)^{1/2} $.
$\mu_t$ is  the   shear modulus.  A WC of classical point 
particles\cite{bm} 
has shear modulus  $\mu_{t,cl}=0.245 e^2 n^{3/2}/ 4\pi 
\epsilon_0\epsilon$,  where we take  $\epsilon=12.8$ the GaAs host dielectric 
constant.     $\omega_0$ is gotten from \mbox{$f_{pk}$} data using the well-known 
formula\cite{flr} for $\omega_0\ll \omega_c$ to take care of the effect 
of the Lorentz force on the dynamics, $\omega_0=(2\pi f_{pk} \omega_c)^{1/2}$.  With 
the classical shear modulus $\mu_{t,cl}$ we then get 
$L =
(2\pi \mu_{t,cl}/neB f_{pk})^{1/2}$.

Figure~4 shows plots of the calculated $L$ vs $\nu$ for all the $n$'s 
we measured.  Panels a and b of the figure show the same data,  with 
successive curves offset for clarity only in panel a.  Most striking 
are the data for  the smallest $n$, $1.8\times 10^{10}$ cm $^{-2}$, which 
 fit a straight line  over the {\em entire range} of 
measurement of the resonance, over a factor of 5 in $\nu$.  The 
fit is accurate to within the instrumental errors in $L$, which we estimate
  as about $\pm 0.02\ \mu$m, propagating mainly from error  in \mbox{$f_{pk}$}.   
The  plots for all the  other  $n$ tend  to curve upwards for larger
$\nu$, as 1/5 is approached.   For the lower $n$'s, the linear behavior 
holds over substantial ranges of $\nu$ and 
is quantitatively demonstrated.   

The straight line regions of the $L$ vs $\nu$ plots 
extend out to extremely low $\nu$, so it is natural to interpret the 
linear behavior as an effect of electron-disorder interaction rather than 
interelectron wavefunction overlap. Particularly   
if it is due to interface roughness\cite{fertig}, the 
disorder potential can vary on extremely 
small length scales,  down to about the lattice constant of GaAs.
 The small lengthscales associated with 
the disorder would make the   effects of electron-disorder interaction 
saturate less easily at high $B$   than interelectron wavefunction overlap effects.  

Because of the clear
linear behavior   for the lowest few $n$'s,
we fit the low $\nu$ tails of all the  $L$ vs $\nu$ plots to 
$L=\beta \nu+L_\infty$, where    $\beta$ and 
$L_\infty$ are the fit parameters, and $L_\infty$ is the domain size 
extrapolated to   infinite $B$. The least squares 
 fits are shown as heavy lines in Figure~4a, and extend  
 over the points that were included in the fits.  
  The inset to 
 Figure~4b shows  the   parameters of the least squares fits.
 The curves appear irregular for larger $n$, and  are more reliable   
  for small $n$,   where the   linear ranges of $L$ vs $\nu$ are 
  substantial. For $n=1.8\times 10^{10}$ cm$^{-2}$, 
$L_\infty\approx 0.17\ \mu\mbox{m}\approx 2.1 a$.\cite{elasticok}

 In the weak pinning energy limit, random pinning  
 models\cite{fertig}-\cite{fogler}   predict  
  $L\sim n$, so that $L$ increases when the WC is stiffer. $L\sim n$ 
  may explain why  the low $n$, low $\nu$  data of Figure~4b group so closely when plotted 
  against  the Landau filling, which contains the density as well as 
  the magnetic length, $\nu=2\pi n l_B^2$.   $L_\infty$ vs $n$   may give 
 some insight into the nature of the pinning in the limit where 
 electrons look like point particles.  In the inset to Figure~4b,    $L_\infty$ 
  vs $n$ clearly increases, as expected for weak pinning. For  $n\le 4.2\times 10^{10}$ cm$^{-2}$,
  $L_\infty$ vs $n$ in the  inset to that figure is roughly  linear, 
  with a fit resulting in a slope of $ 2.3\times 10^{-22}$ m$^3$, and intercept of 0.13 $\mu$m.

We interpret the upward curvature of  the $L$ vs $\nu$  data 
as 1/5 is approached
(except for $n=1.8\times 10^{10}$ cm$^{-2}$) 
as effects of correlations related to the FQHE, which  
require interelectron overlap.    One possibility  is 
that   $\mu_t$ is decreasing as 1/5 is approached 
due to these correlations, as predicted in a recent theory\cite{narevich} of a WC made up of 
composite fermions.    $L$ presented in  Figure~4 is calculated using  the classical shear modulus, $L\sim 
\mu_{t,cl}^{1/2}$, and would   {\em overestimate} the true 
domain size, $\tilde{L}$,  if the true  shear modulus $\mu_t$ is less than 
$ \mu_{t,cl}$.  This overestimation in  the $L$ data would 
result in upward curvature of the plots, even if  the underlying 
$\tilde{L}$ vs $\nu$
  were  straight lines, like the  $n=1.8\times 10^{10}$ cm$^{-2}$
plot in Figure~4.  We interpret the reduction in the curvature at low $n$ 
 as due to the suppression of the FQH type correlations, owing to the 
 increased importance of disorder relative to electron-electron 
 interaction at lower $n$.  This is consistent with  Figure~1b,
 in which the 1/5 feature is suppressed at  lower $n$.

In summary, systematic  studies of the microwave resonance for many 
$n$'s show  maxima in $f_{pk}$ vs $\nu$; these  
curves   are greatly simplified on converting \mbox{$f_{pk}$} into a domain size $L$.  
$L$ vs $\nu$   fits  a straight line for low $n$ and low $\nu$, but 
curves upward as $\nu=1/5 $ is approached. The straight line behavior 
is interpreted as an effect of interplay between electron 
wavefunctions and disorder features; the upward curvature is 
interpreted as an effect of FQH-type correlations, possibly associated 
with softening of the WC. 

The authors wish to thank R. Chitra, H. Fertig, M. Fogler and D. Huse for valuable 
discussions.  This work is supported by the Air Force Office of Scientific Research, 
and the National Science Foundation.

\begin{figure}[t] 
  \includegraphics[scale=0.45]{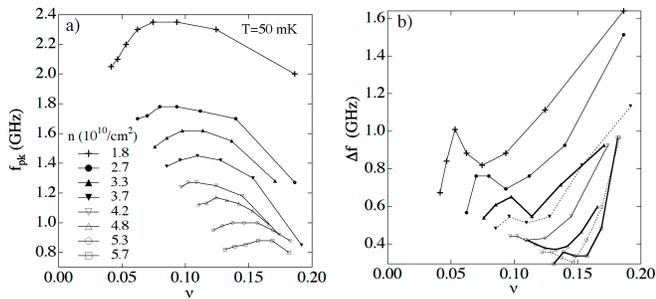}
 \caption{
 {\em a. \ \ }Peak frequency $f_{pk}$ vs Landau filling $\nu$ for   different carrier 
 densities, $n$. \ \ \ \ {\em  b.\  \ } Full width at half maximum 
 linewidth, $\Delta f$, vs $\nu$. Symbols  are same as in 
 panel a.  \vspace*{-7pt}  }       
 \end{figure}
\begin{figure}[t] 
 \includegraphics[scale=0.43]{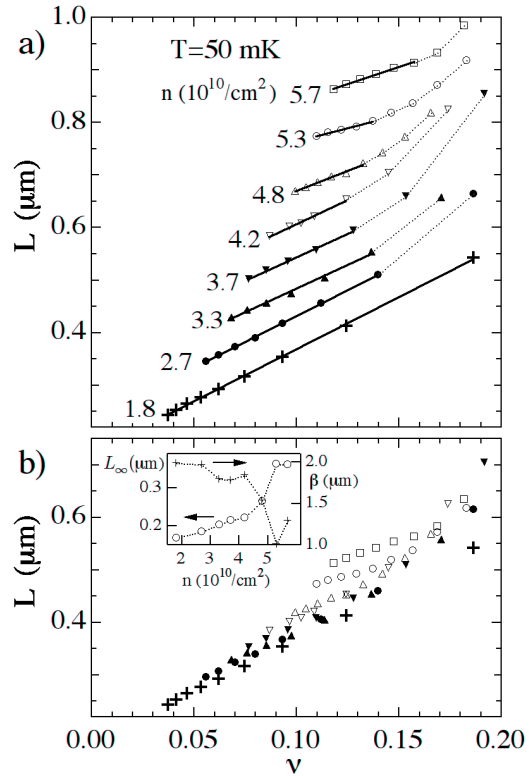}
 \caption{ Calculated domain sizes $L$, vs Landau filling $\nu$ for various carrier 
 densities $n$.   {\em a. \ \ } Successive curves are offset from each 
 other by 0.05 $\mu$ m. Curves are labeled 
 with $n$ ($10^{10}$ cm$^{-2}$). Heavy lines are  linear least squares 
 fits to $L=\beta \nu + L_\infty$, extending over ranges of $\nu$ that were fit. 
 {b. \ \ } Same data as in panel {\em 
 a}, but  not offset.  Symbols for the various $n$ are the same as in 
 {\em a}.  Inset shows  fit parameters, $\beta$ and $L_\infty$, vs $n$. 
  }
 \end{figure}


\begin{thebibliography}{}
\bibitem{bzeronote}In the   $B=0$ equivalent, the high $B$ shear modulus and 
electron-impurity interactions are retained: only the Lorentz force 
that mixes the elastic and cyclotron response is turned off.
\bibitem{elasticok}  If $\xi\ll a$, then $L_a\gg L$ 
 is possible and    $L\sim a$ obtained here does not conflict 
 with  the use of an elastic model to obtain $L$.
\end{thebibliography}
\end{document}